\documentclass[usenatbib]{mn2e}

\usepackage{times}
\usepackage{graphicx}
\usepackage{amssymb}

\begin{document}

\title{Portrait of Theobalda as a Young Asteroid Family}

\author[Bojan\ Novakovi\'c]{Bojan~Novakovi\'c
\thanks{E-mail: bojan@matf.bg.ac.rs} \\
Department of Astronomy, Faculty of Mathematics, University of Belgrade,
Studentski trg 16, 11000 Belgrade, Serbia}

\maketitle \begin{abstract} 
The (778)~Theobalda asteroid family attracted little attention so far, but our study shows that it is important in several aspects.
In this paper we investigate the origin and evolution of Theobalda family.
Firstly, we identify the family as a statistically relevant group in the space of synthetic proper elements. Using the hierarchical clustering method
and adopted cut-off velocity of $d_{cutoff}=85$~ms$^{-1}$ we found that Theobalda family currently consists of 128 members.
This family is located in the outer belt, near proper semi-major axis $a_{p}$~$\approx$~3.175~au. This region is crossed by
several three-body mean motion resonances which give rise to significant chaotic zones. Consequently, the majority of family members 
reside on chaotic orbits. Using two independent methods, chaotic chronology and backward integration, we found Theobalda family
to be only 6.9~$\pm$~2.3~Myr old. We have also estimated, that the family was likely produced 
by the cratering impact on a parent body of diameter $D_{PB}$~$\approx78\pm9$~km.
\end{abstract}



\section{Introduction}
\label{s:intro}

Asteroid families are believed to originate by a catastrophic disruption of large asteroids.
To identify an asteroid family, one looks for clusters of asteroids 
in the space of proper orbital elements \citep{MilKne90,MilKne94}: the proper
semi-major axis ($a_{p}$), proper eccentricity ($e_{p}$) and proper inclination
($I_{p}$). The orbital elements describe the size, shape and tilt of orbits.
Proper orbital elements, being more constant over time than
instantaneous ones, provide a dynamical criterion of whether or not a group 
of bodies has a common ancestor.
Up to now, ejecta from a few tens of major collisions have been discovered in the main belt \citep[e.g.][]{zappala94,Diniz05}.

The size and velocity distributions of the
family members provide important constraints for testing our
understanding of the break-up process, but erosion and dynamical 
evolution of the orbits over time can alter the original
signature of the collision. It is nowadays well known that the kinematical structures of 
the asteroid families evolved over the time, with respect to the original 
post-impact situations, due to chaotic diffusion, gravitational and non-gravitational 
perturbations \citep{MilFar94,bottke2001,carruba03,delloro04}. 
These mechanisms changed the original shapes of the families produced in 
collisions, and consequently complicated physical studies of high-velocity collisions.

Unfortunately, most of the observed asteroid families are old enough 
(older than 100 Myr \citep{nes2006}) to be
substantially eroded and dispersed. On the other hand, young asteroid families 
(younger than 10~Myr) such as Karin, Veritas and Iannini \citep{nes2002,nes2003} 
or even very young families (younger than 1~Myr) such as
Datura, Emilkowalski, 1992YC2, and Lucascavin \citep{nesvok2006},
suffer little erosion during the period of time after a breakup event.
Thus, they provide a unique opportunity to study a collisional outcome almost
unaffected by orbit evolution.

In this paper we study Theobalda asteroid family. We present its basic properties including
the identification of its membership, and the study of cumulative absolute magnitude distribution of the family members. 
Moreover, the diameter of the parent body has been estimated. We studied in detail the dynamical characteristics
in the region occupied by the Theobalda asteroid family and analyzed the role of the dynamics in shaping the family . 

As was noted by \citet{nov09} this family is a very good candidate
to estimate its age by the method of chaotic chronology (MCC). In order to apply MCC 
the family has to be located in the region of the main asteroid belt where
diffusion takes place. Also, it is necessary that diffusion is fast enough to cause measurable effects, 
but slow enough so that most of the family members are still forming a robust family structure. 
As we show, it turns out that Theobalda family is an excellent case in this respect. Given that, as our main result 
we have estimated the age of the family. Using two different methods, MCC \citep{menios07,nov09} and backward integrations \citep{nes2002},
we estimate the age of the family to be 6.9~$\pm$~2.3~Myr. Thus, we establish it as another young asteroid family.

The paper is organized as follows: In Section~\ref{s:basic} we present the basic properties
of the Theobalda family. We use the hierarchical clustering method (HCM), proposed by
\citet{zappala90}, to identify family members. Next, the cumulative absolute magnitude distribution of
identified family members is discussed, and the size of parent body is estimated. 
The dynamical characteristics of the region occupied by Theobalda family 
are presented and discussed, and main mean motion and secular resonances, in that region, 
identified. The dynamical stability of the family members is analyzed, in particular
the stability of the largest member of the family (778)~Theobalda. 
In Section~\ref{s:age} we estimate the age of the family. 
This is performed firstly by using the backward integration method, and then by using the method of chaotic chronology. 
The good agreement between these two results indicates a reliable age determination. Finally, in Section~\ref{s:conclusions}
we summarize our results, discus some possibly interesting relations to other works,
and draw our conclusions.

\section{Theobalda family: the basic facts}
\label{s:basic}

This asteroid family has attracted little attention so far, mainly because the number of
asteroids associated with it was relatively small. However, the situation is different at present,
and, as we will show later, Theobalda family now has over 100 known members. This number
is large enough that the family characteristics can be reliably determined.

\subsection{Identification of the family members}
\label{ss:hcm}

\begin{figure}
\begin{center}
\includegraphics[height=0.33\textheight,angle=-90]{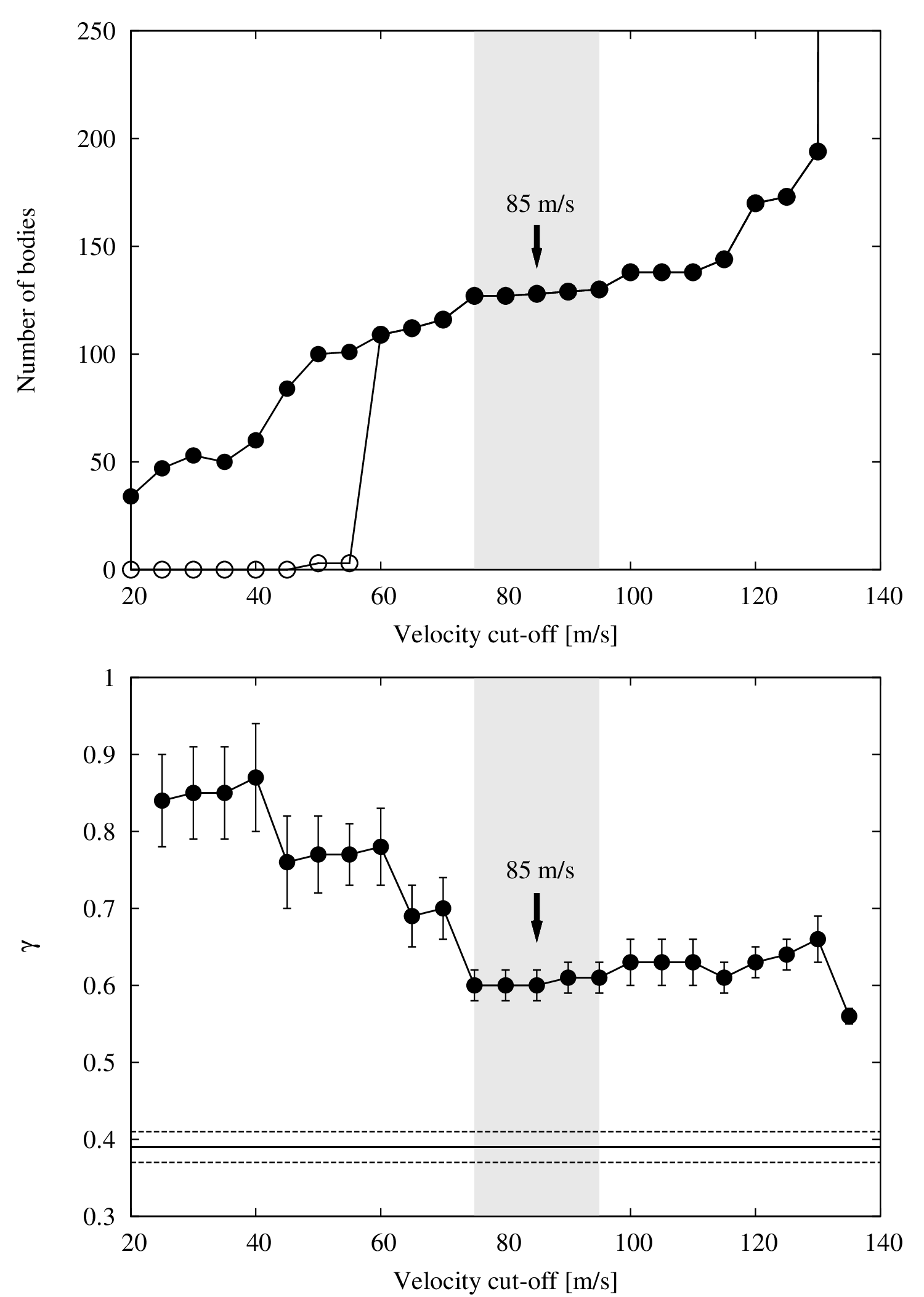}
\end{center}
\caption[]{Number of asteroids associated with Theobalda family (top) and a power-law index $\gamma$ of 
the cumulative magnitude distribution in the range H$\in$[13-15] (bottom) as a function of the cut-off 
distance ($d_{cutoff}$). Note that in both cases the respective values are nearly constant for 
$d_{cutoff}$~$\in$[75,95]~ms$^{-1}$. In the bottom panel horizontal solid line and two dashed lines 
represent background level and its error-bars respectively. In the top panel two obvious critical 
values of $d_{cutoff}$ are: (i) $60$~ms$^{-1}$, when group around (778)~Theobalda merge with the group 
around (84892) $2003QD_{79}$, and (ii) $130$~ms$^{-1}$, when the family starts to merge with the local 
background population. However, the number of asteroids associated with the family is constant for 
$d_{cutoff}$~$\in$[75,95]~ms$^{-1}$. For the nominal family we chose $d_{cutoff}=85$~ms$^{-1}$ 
(see text for additional explanation).}
\label{f:cutoff}
\end{figure}

The identification of family members is the first step in our study of the Theobalda family.
This is done by applying the HCM to the catalog of synthetic proper elements of numbered asteroids
\citep{KneMil00,synthpro2} from AstDys\footnote{http://hamilton.dm.unipi.it/astdys/} 
(database as of October 2009). The HCM requires 
that distances among the family members, in the proper elements space, are less than the so 
called \textit{cut-off} distance ($d_{cutoff}$), which has dimension of velocity. As the cut-off 
distance is a free parameter of HCM, we tested different values ranging from $20$ to 
$135$$~ms^{-1}$. Also, we apply HCM using two different \textit{central objects}:
(i) (778)~$Theobalda$ which has a chaotic orbit, 
and (ii) (84892) $2003QD_{79}$ which is on
the relatively stable orbit. The results are shown in the top panel of Fig.~\ref{f:cutoff}. 
The HCM identified the family around (778)~Theobalda for $d_{cutoff} \geq 60$$ms^{-1}$, while around (84892) $2003QD_{79}$ family
exists even for lowest tested value of $d_{cutoff} = 20$$~ms^{-1}$. For $d_{cutoff} \geq 60$$~ms^{-1}$
resulting family is the same. This suggests that (778)~Theobalda probably has been displaced from
its original position due to the chaotic diffusion.\footnote{Note, that this is very similar
to Veritas family, and situation with the largest member of this family (490) Veritas \citep{menios07}.}
In the bottom panel of Fig.~\ref{f:cutoff} the best-fit power-law index $\gamma$ of the form N($<$H)~$\varpropto$~$10^{\gamma H}$ 
of the cumulative absolute magnitude (H) distribution in the range H$\in$[13-15], as a function of cut-off distance ($d_{cutoff}$), 
is shown\footnote{Instead of the index $\gamma$, the exponent of the cumulative distribution can be obtained
in terms of diameters rather than absolute magnitudes \citep{delloro07}. However, as we do not know albedos 
for most of the asteroids, necessary to convert from absolute magnitudes to diameters, we chose 
to work with $\gamma$.}.
For  $d_{cutoff}$~$\in$[75,115]~ms$^{-1}$ the number of asteroids as well as index $\gamma$ are nearly constant, 
and probably each value from this interval can be safely used to identified family members by HCM.\footnote{Usually one adopts the value
of $d_{cutoff}$ that corresponds to the center of the interval over which the index $\gamma$ is constant \citep{vok06b}.}
We adopted value of $d_{cutoff}=85$~ms$^{-1}$ to identify \textit{nominal} family.
For this value of $d_{cutoff}$, HCM linked 128 asteroids to Theobalda family.
There are two main reasons for our choice. The first one is that
this value of velocity cutoff corresponds to the center of the \textit{plateau} which can be seen in Fig.~\ref{f:cutoff}.
The second reason is very good agreement between the ages of family estimated applying MCC to two different groups of 
family members. We will explain this in more detail in Section \ref{s:age}.

Note that the values of $\gamma$ for family members are always larger than the value of the same index calculated for background 
asteroids. This is the first indication that the family is relatively young. On the other hand, \citet{parker08} 
estimated $\gamma$=0.44 for H$\in$[13.0-15.5]. This value is much lower than ours, as we found $\gamma$=0.60$\pm$0.02 for the nominal family, 
and very close to the value that we found for background population in the region of Theobalda family. 
Probably, \citet{parker08} underestimated this value due to the observational incompleteness, as they worked in the 
range H$\in$[13-15.5] and used smaller dataset for which Sloan Digital Sky Survey (SDSS) colors were available. Although they linked 
100 asteroids with the family, a significant number of these asteroids are probably interlopers.

\subsection{Size of the parent body}
\label{ss:pb}

\begin{figure}
\begin{center}
\includegraphics[width=0.33\textwidth,angle=-90]{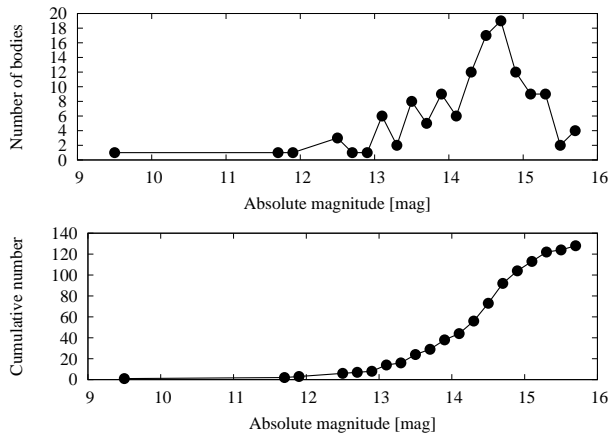}
\end{center}
\caption[]{Distribution (top) and cumulative distribution (bottom) of the members of Theobalda asteroid family 
as a function of the absolute magnitude.}
\label{f:sfd_bw}
\end{figure}

\begin{figure}
\begin{center}
\includegraphics[width=0.33\textwidth,angle=-90]{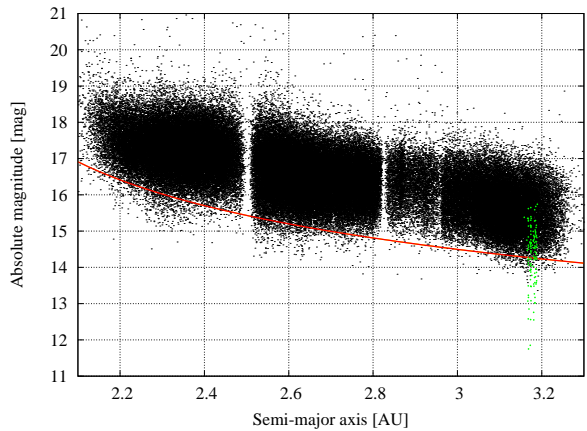}
\end{center}
\caption[]{The distribution of multi-opposition asteroids, in the (a,H) plane, from AstDys web site (database as of October 2009). 
The catalog consists mostly of the objects discovered more recently than objects included in the catalog of synthetic proper elements. 
Thus, it provides good opportunity to estimate completeness limit of the catalog of synthetic proper elements, which is marked by a red line.
The green dots represent the members of Theobalda family.}
\label{f:MB_aH}
\end{figure}

To estimate the diameter of the parent body ($D_{PB}$) of the Theobalda family, it is
necessary to account for small and still undiscovered family members. In general, 
data set on asteroids below $H$=15~mag is basically complete \citep{gladman09}. However, as we 
are dealing with the family at the edge of outer belt, the family members are 
the C-type asteroids which are several times darker 
than the S-type asteroids, and, since we are using catalog of synthetic proper elements which does not
include all known asteroids, the completeness limit for our sample has to be analyzed.
An indication about the completeness limit can be obtained by simply looking at Fig.~\ref{f:sfd_bw} for the value of $H$ 
where slopes of two distribution curves change. This is approximately at about $H$=14.5~mag.
Somewhat better estimation of completeness limit can be inferred using the catalog of asteroids, which are not included in the
catalog of synthetic proper elements we deal with, i.e. the catalog of multi-opposition objects maintained at AstDys web site. 
As about 99 per cent of multi-opposition objects, with osculating semi-major axes in the range [3.15,3.20]~au, 
have $H\geq$14.2~mag (see Fig.~\ref{f:MB_aH}), we assume that the catalog of synthetic proper elements 
of asteroids in this region is complete up to $H$=14.2~mag.

In order to overcome the problem of observational incompleteness, using the size-frequency 
distribution (SFD) of the main belt asteroids estimated by \citet{gladman09}\footnote{The asteroid size distribution
at diameters D$<$10~km is still poorly known. Various models and extrapolations yield 
very different estimates of the number of km-sized and smaller main belt asteroids.
However, other possible estimates \citep[e.g.][]{ivezic01,ted05,wiegert07,yoshida07}
would not affect our results significantly.} and the fact that SFD of asteroid families 
is considered to be somewhat shallower than that of the background \citep{morby2003} we added some fictitious bodies with
$H\in$[14.2,17.0] to the family. This was performed in such a way to make SFD of ''extended'' family (real asteroids + fictitious objects),
to be somewhat shallower than obtained SFD for background asteroids. More precisely, in order to be able to estimate the uncertainty of our approach,
we generated 100 different sets of fictitious objects and then estimated the size of the parent body from each of the 100 sets.

$D_{PB}$ corresponds to a spherical body with volume 
equal to the estimated total volume of all the family members, including these
added members with $H\in$[14.2,17.0]. Next, we assume all family members have the same 
geometric albedo ($p_{v}$) as (778)~Theobalda, that is $p_{v}$=0.0589 according to \citet{ted02}.
Having the values of $H$ and $p_{v}$, the radius $R$ of a body can be estimated, using
the relation \citep[e.g.][]{bowell1989}
\begin{equation}
 R~{\rm (km)} = 1329~ \frac{10^{\frac{-H}{5}}}{2\, \sqrt{p_{v}}}
\end{equation}
This allows us to infer that the diameter\footnote{Probably the better way to estimate the size of the parent body is the one
proposed by \citet{durda07}. However, we were unable to find appropriate match with SFDs published in that paper using simple
visual comparison of plots.} of the parent body was $D_{PB}$~$\approx78\pm9$~km. The estimated uncertainty accounts for uncertainties in
albedos, absolute magnitudes and SFD. Also it takes into account the dependence on the HCM cut-off. However, the real uncertainty is somewhat larger, e.g. because of the possible interlopers.
According to our estimate, the largest remnant (778)~Theobalda contains 87 per cent of the mass of the parent body.
Although this should be considered as an upper bound, because it does not account for small family members with $H>$~17, our result suggest
that Theobalda family was produced by a cratering impact.
The typical density ($\rho$) for C-type asteroids is $\rho=1500~$kg~m$^{-3}$ 
\citep[e.g.][]{broz05}. 
Given that, the escape velocity\footnote{Compensating for collective effects in the cloud of dispersing fragments, 
$V_{esc}$~=~1.64~$\times~GM/R$, where $GM$ is the product of the 
gravitational constant and the parent body mass, $R$ is radius of the parent body, while 1.64 
is an empirical factor \citep{petfar1993}.} from Theobalda family parent body was $V_{esc}$$\approx$~32~ms$^{-1}$.

\subsection{Dynamical characteristics}
\label{ss:dynamics}

The dynamics in the region of the phase space occupied by Theobalda family members is much
like in the case of the Veritas family because two families stretch over the similar range of
the proper semi-major axes. The dynamics in the region of Veritas family is very well 
studied \citep[see e.g.][]{MilFar94,KnePav02,nes2003,menios07}. Therefore, here we will
focus only on some differences between dynamics in the regions occupied by two families.
The differences arise from the fact that proper eccentricities of Theobalda family members 
($e_{p}$$\approx$0.25) are significantly higher than those of Veritas family ($e_{p}$$\approx$0.06).
Also, the proper inclinations are by about $5^{o}$ higher. These make Theobalda family members
even more strongly chaotic than Veritas family members. 

\begin{figure}
\begin{center}
\includegraphics[width=0.33\textwidth,angle=-90]{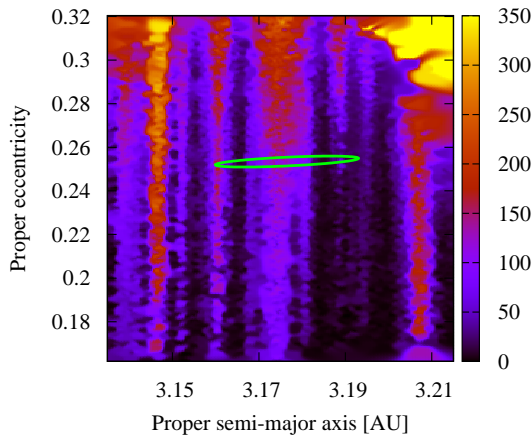}
\end{center}
\caption[]{The dynamical structure of the region occupied by Theobalda family along with surrounding area. 
The color scale codes Lyapounov Characteristic Exponents (in the units of $10^{-6}$~$yr^{-1}$) for 10,000 test particles. 
The ellipse represents assumed positions of the Theobalda family members immediately after break-up.}
\label{f:LCE}
\end{figure}

\begin{figure}
\begin{center}
\includegraphics[width=0.33\textwidth,angle=-90]{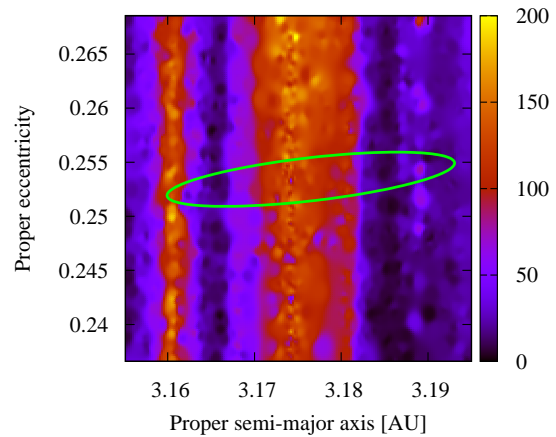}
\end{center}
\caption[]{The same as in Fig.~\ref{f:LCE} but for LCEs of the real family members. 
The linear interpolation is used in order to cover the complete region shown in the figure.}
\label{f:tlyap}
\end{figure}

In Fig.~\ref{f:LCE} the Lyapounov Characteristic 
Exponents (LCEs), as a function of proper semi-major axis and eccentricity, in the region occupied 
by Theobalda family members, as well as in the surrounding area are shown.
Most of the orbits in that region are unstable even for comparatively low values of proper eccentricity, 
while for eccentricity above 0.3 almost all chaotic zones are connected forming a wide \textit{chaotic sea} with a fast diffusion therein. 
The chaos is also dominating in the region 
where Theobalda family is located (inside or close to the equivelocity ellipse). This can be better appreciated from
Fig.~\ref{f:tlyap} where LCEs of Theobalda family members are shown. The vertical strip of the largest at $a_{p}$$\approx$3.174~au
values of LCEs is associated with (5, -2, -2) three-body\footnote{All three-body mean motion resonances \citep{nesmor98} discussed 
in this paper are among Jupiter, Saturn and asteroid.} MMR, but it seems that this chaotic zone includes 
(3, 3, -2) and (7, -7, -2) three-body MMRs as well. Most of the bodies have LCE~$\geq$1$\times$$10^{-4}$~$yr^{-1}$, 
which corresponds to the Lyapounov times $T_{lyap}$ $\leq$ 10,000~yr and these bodies are probably in the 
so-called \textit{Chirikov regime} \citep{guzzo2002,morby02}.

\begin{figure}
\begin{center}
\includegraphics[height=0.33\textheight,angle=-90]{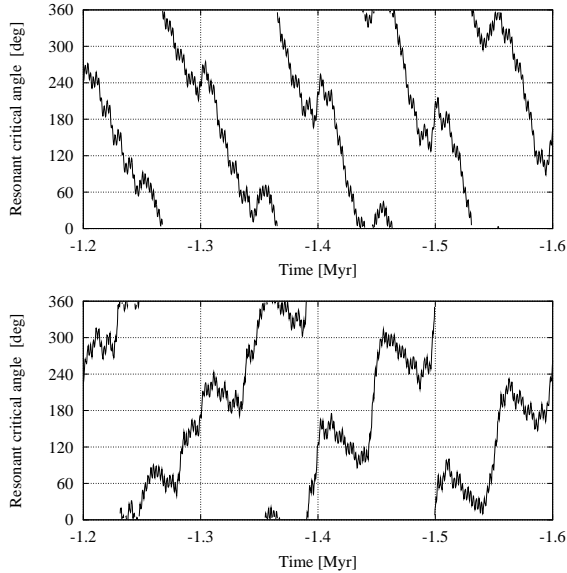}
\end{center}
\caption[]{Time evolution of the critical angles
$\sigma_{1}$=$\varpi$+$\Omega$-$\varpi_{5}$-$\Omega_{6}$ (top) and $\sigma_{2}$=$\varpi$+$\Omega$-$\varpi_{6}$-$\Omega_{6}$ (bottom) 
of the secular resonances for period from 1.2~Myr to 1.6~Myr in the past. 
The short episodes of trapping inside the resonance seem possible, but several events of resonance crossing are clearly visible.
Reversal of direction of circulation is related to periods when the orbit interact with these secular resonances.}
\label{f:secular_resonances}
\end{figure}

Using the proper frequencies $g$ (average rate of the perihelion longitude
̟$\varpi$) and $s$ (average rate of the node longitude $\Omega$)\footnote{The secular
frequencies of Jupiter ($g_{5}$) and Saturn ($g_{6}$,$s_{6}$) are taken from \citet{nobili1989}.} we found that
Theobalda family region is also crossed by two secular resonances $g+s-g_{5}-s_{6}$ 
and $g+s-g_{6}-s_{6}$ (Fig.~\ref{f:secular_resonances}). Both are of the order 4, i.e. they arise 
from the perturbing terms of degree of at least 4 in eccentricity and inclination \citep{MilKne90,kne1991}.
However, we did not find evidence (see Section \ref{sss:diffusion}) that these resonances 
increase diffusion speed. Probably this is because these resonances are effective only in narrow bands within 
the Theobalda family\footnote{The secular resonance $g+s-g_{6}-s_{6}$ has much more influence on dynamics of 
family members in the case of Padua family \citep[see][]{carruba09}.}. Consequently, some of the family members, 
during their secular cycles, might be temporally trapped in one 
or both secular resonances, but most of the time these asteroids are outside the secular resonances.
In Fig.~\ref{f:secular_resonances} the time evolution of the critical angles 
$\sigma_{1}$=$\varpi$+$\Omega$-$\varpi_{5}$-$\Omega_{6}$, and $\sigma_{2}$=$\varpi$+$\Omega$-$\varpi_{6}$-$\Omega_{6}$, for asteroid 
(778)~Theobalda are shown. This asteroid might be temporally trapped in both secular resonances. 
Although the short episodes of ''libration'' are visible, these events may be related to resonance crossing rather then to the resonance trapping.
Most of the time both critical arguments circulate.

\begin{figure}
\begin{center}
\includegraphics[height=0.33\textheight,angle=-90]{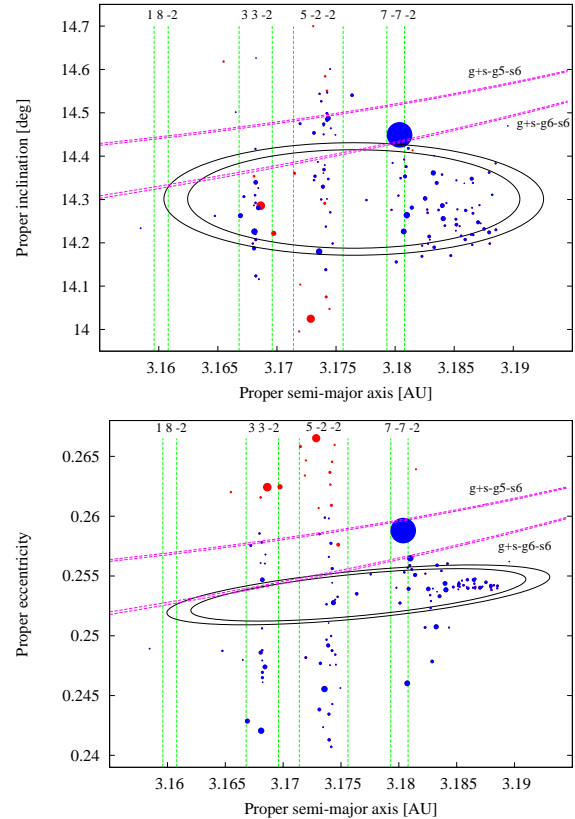}
\end{center}
\caption[]{Distribution of the known Theobalda family members in the
($a_{p}$,$I_{p}$) plane (top) and ($a_{p}$,$e_{p}$) plane (bottom). The
superimposed ellipses represent equivelocity curves, computed according to the
equations of Gauss \citep[e.g.][]{morby95}, for a velocities of
$v=35$~ms$^{-1}$ (inner) and $v=40$~ms$^{-1}$ (outer), true anomaly $f=85^{\circ}$ 
and argument of pericentre $\omega=95^{\circ}$. The blue points represent family members identified for $d_{cutoff}=65$~ms$^{-1}$, 
while red points represent additional family members identified for $d_{cutoff}=85$~ms$^{-1}$. 
The size of each point corresponds to the diameter of the body. The green dashed lines mark approximately borders of three-body MMRs, 
while pink dashed lines show locations of secular resonances.}
\label{f:dis_aei}
\end{figure}

In Fig.~\ref{f:dis_aei} distributions of family members, as identified by HCM, are shown
along with the positions of main mean motion and secular resonances.
Obviously, the structure of the family is a result of dynamical mechanisms at work,
which are mainly controlled by MMRs. The largest spread of family members, in both ($a_{p}$,$e_{p}$)
and ($a_{p}$,$I_{p}$) planes, is associated to (5,~-2,~-2) resonance. Somewhat smaller spread is
observable in the (3,~3,~-2) resonance, while (7,~-7,~-2) resonance caused only small diffusion of
asteroids. This agree very well with obtained values of LCEs (see Fig.~\ref{f:tlyap}).

\begin{figure}
\begin{center}
\includegraphics[width=0.33\textwidth,angle=-90]{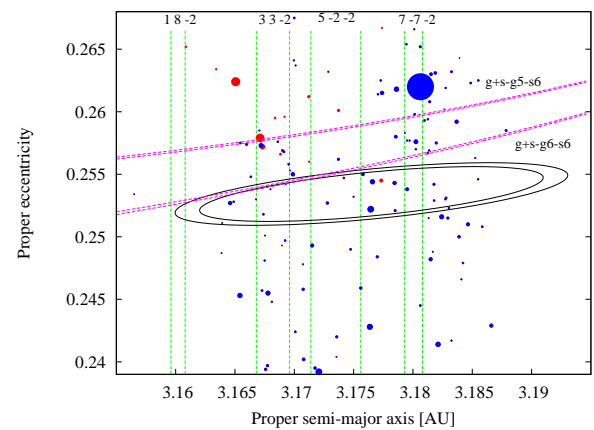}
\end{center}
\caption[]{The same as in Fig.~\ref{f:dis_aei} (bottom), but for analytical proper elements.}
\label{f:analytical}
\end{figure}

\begin{figure}
\begin{center}
\includegraphics[width=0.33\textwidth,angle=-90]{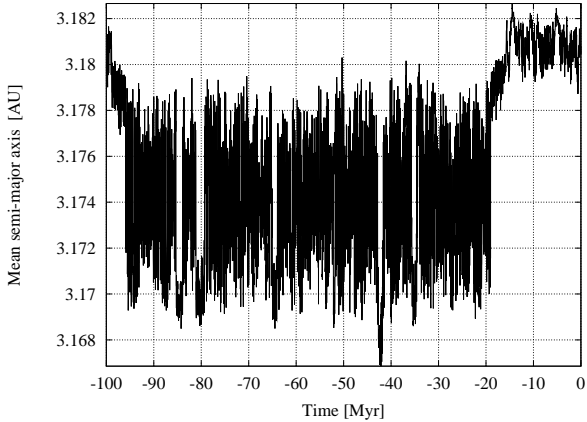}
\end{center}
\caption[]{Time evolution for 100~Myr back-in-time of the mean semi-major axis for asteroid (778)~Theobalda. 
For the first about 15~Myr of integration it resides in (7,~-7,~-2) resonance, but then, it switches to (5,~-2,~-2) resonance 
where it spends most of the time covered by integrations. Also, it exhibits short episode of trapping inside (3,~3,~-2) resonance
at about 42~Myr.}
\label{f:778}
\end{figure}

It is interesting to note that there are gaps (without family members) between the (3,~3,~-2) and (5,~-2,~-2) resonances,
as well as between the (5,~-2,~-2) and (7,~-7,~-2) resonances. We suggest that this is another confirmation that all these
three resonances are connected and make one wide chaotic zone. Because of this, all asteroids from $a_{p}$$\approx$3.167~au 
to $a_{p}$$\approx$3.181~au reside in one of these three resonances. The asteroids can switch from one resonances to another
(see Fig.~\ref{f:778}), but on the time scale of a few Myr this is a rare event, so that each asteroid spends most of 
the time in one of the resonances. As a result, due to the some uncertainty in the procedure of computation of synthetic proper elements
for resonant asteroids, i.e. averaging does not work well, all bodies appear to be located in (or close to) the center of one of
the resonances. This can be verified by using the analytical proper elements\footnote{We did not use analytical proper elements, 
in our other analysis throughout the paper, because they are not enough
accurate in this high eccentricity region. This can be appreciated comparing the distributions of Theobalda family members with $a_{p}$$\geq$3.183~au,
shown for two different kinds of proper elements. Obvious grouping of the regular members, which is clearly visible in the space of synthetic proper elements 
(Fig.~\ref{f:dis_aei}), disappears in the space of analytical proper elements (Fig.~\ref{f:analytical}).} of \citet{MilKne90}. 
These elements are calculated by means of analytical theory
based on the series development of the perturbing Hamiltonian, and which does not include averaging. 
The distribution of Theobalda family members in the space of analytical proper elements (in the $a_{p},e_{p}$ plane) is
shown in Fig.~\ref{f:analytical}. The shown distribution is roughly random and without gaps in terms of proper semi-major axis, 
what confirms our claim that the gaps appear due to the averaging procedure. Moreover, it means that switching from one resonance to another must be a rare event, 
but, the fact that not all of the asteroids are located in the center of one of the resonances, is another evidence that resonance switching is possible,
i.e. these three resonances are connected.

The position of the largest remnant, asteroid (778)~Theobalda, is not close to the
center of the family. This is evident also in ($a_{p}$,$I_{p}$) 
plane, but it is more obvious in ($a_{p}$,$e_{p}$) plane. As we already mentioned above, this 
asteroid has probably been displaced from its original position due
to the some dynamical mechanisms. It is located close to or inside the (7,~-7,~-2) three-body MMR,
which might be responsible for its relatively high proper eccentricity. However, its proper semi-major axis is 
also larger than that of the center of family, and this could not be explained by
(7,~-7,~-2) resonance. Because of that, we investigate dynamics of this asteroid in more detail.
The orbit of (778)~Theobalda is propagated for 100~Myr back in time\footnote{All 
integrations presented in this paper are performed using the public domain
ORBIT9 integrator embedded in the multipurpose OrbFit package (http://hamilton.dm.unipi.it/astdys/), 
and dynamical model that includes the four major planets 
(from Jupiter to Neptune) as perturbing bodies. The indirect
effect of the inner planets is accounted for by applying a barycentric
correction to the initial conditions.}. As we will show later, the family is about 6-7~Myr old.
Why than it is meaningful to integrate 100~Myr? The answer is hidden in the chaotic motion of this
asteroid. As we know, chaos is not predictable on the time scales of several times the inverse of LCE,
which is in the case of (778)~Theobalda $\approx$8,000~yr. All that we can achieve is to show what kind of 
behavior (i.e. motion) is possible. In this respect, our 100~Myr long integrations are equivalent
to many shorter integrations with slightly different initial conditions. Similar technique was used
by \citet{laskar94} to study stability of the Solar system.

Fig.~\ref{f:778} shows 100~Myr of the back-in-time 
evolution of the mean semi-major axis of asteroid (778)~Theobalda. Initially, its semi-major axis oscillates around 
3.181~au (close to (7,~-7,~-2) resonance), but after about 15~Myr (in the past) the value of the semi-major axis drops to $\approx$3.174~au.
Around this time it actually switches from (7,~-7,~-2) to (5,~-2,~-2) resonance. There is another switch of the resonance about
42~Myr, when the asteroid is temporally trapped in (3, 3, -2) resonance at $\approx$3.168~au. Finally, at about 95~Myr in the past, it 
went back from (5,~-2,~-2) to (7,~-7,~-2) resonance. This is the confirmation that these three resonances are connected at higher 
eccentricities ($e_{p}$~$\geq$~0.25). But again, due to the chaoticity, the behavior of the mean semi-major axis of asteroid 
(778)~Theobalda does not represent quantitatively its real motion, but qualitatively. Still, behavior of its semi-major axis suggests
that chaos may be responsible for displacement of (778)~Theobalda from the center of family in terms of $e_{p}$ and $I_{p}$.
However, if this was the case, this asteroid probably spend some time residing in (5,~-2,~-2) resonance which is strong enough
to increase its eccentricity from $e_{p}$$\approx$0.253 to $e_{p}$$\approx$0.259, on the time scale of several Myr.

Studying the distribution of family members shown in Fig.~\ref{f:dis_aei}, it can be noted that there are no family members located
inside equivelocity ellipses, at $a_{p}$$\approx$3.165~au. Contrary to the gaps between the resonances, the absence of asteroids that belong to the family
in this region cannot be explained by dynamical instability or by ''weakness'' of the procedure of proper elements calculation.
Although, a detail study of this problem is beyond the scope of our work, we believe that this may be related to the impact characteristics
(cratering event), which ''forced'' fragments to be symmetrically distributed around the semi-major axis of the largest fragment (778)~Theobalda.

\section{The age of Theobalda family}
\label{s:age}

\subsection{Backward integration}
\label{ss:backwards}

Backward integration of orbits is very accurate method for family age estimation, which works well with
young families. It is based on the fact that due to the planetary perturbations 
the orientation of orbits in the space changes over time. Consequently, two angles 
that determine the orientation of orbits in space, the longitude of the ascending 
node ($\Omega$) and the longitude of perihelion ($\varpi$), evolve with different 
but nearly constant speeds for individual orbits. After some time this effect
spreads out $\Omega$ and $\varpi$ uniformly around $360^{o}$.
On the other hand, immediately after the disruption of the parent body, the orientations 
of the fragments' orbits must have been nearly the same. Given that, the age of an asteroid
family can be determined by integrating the orbits of the family members backwards,
until the orbital orientation angles cluster around some value. This method was used by
\citet{nes2002,nes2003} to determine the ages of the Karin cluster (5.8$\pm$0.2~Myr) and 
Veritas family (8.3$\pm$0.5~Myr).

\begin{figure}
\begin{center}
\includegraphics[width=0.5\textwidth,angle=-90]{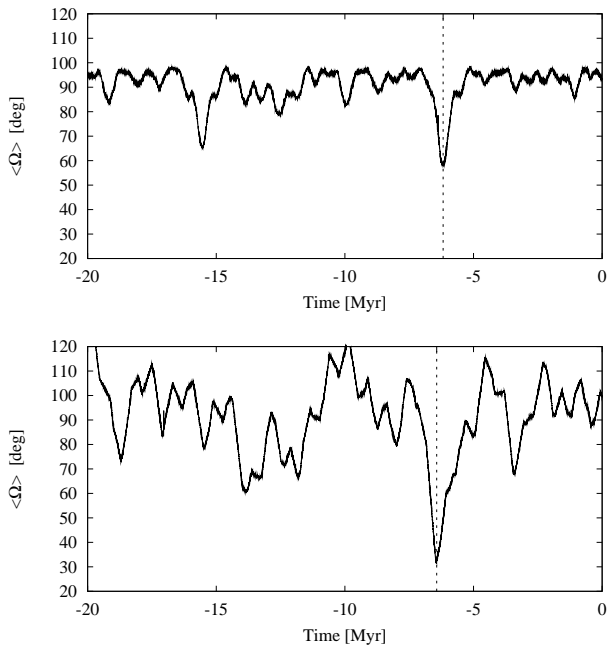}
\end{center}
\caption[]{The average differences in nodal longitudes ($\langle\Delta\Omega\rangle$) for 13 members of Theobalda family, with
regular orbits. The results obtained in a purely gravitational model (top) and with yarko clones (bottom), are shown. 
The most important feature is clustering at about 6.2~Myr ago within $\approx$$58^{o}$. 
This minimum becomes significantly deeper ($\approx$$31^{o}$) with yarko clones, and also slightly shifted to 6.4~Myr ago.}
\label{f:clustering}
\end{figure}

Here we applied Nesvorn\' y et al.'s method to try to estimate the age of Theobalda family.
By integrating the orbits of the Theobalda asteroid family back in time, hopefully, we can
find a conjunction of orbital elements ($\Omega$ and $\varpi$), which occurred only 
immediately after the disruption of the parent body.
This method is, however, limited to groups of objects moving on regular orbits, which, even in that case,
can be accurately track up to 20~Myr in the past.
Similarly as in the case of Veritas family \citep{nes2003,menios07}, only a 
fraction of Theobalda family members satisfies this condition and can be accurately integrated back 
in time. Also, as was pointed out by \citet{nes2003} \citep[see also][]{nes2008} the region
around $a_{p}$=3.175~au is close enough to the 2/1 MMR with Jupiter to undergo fast 
differential evolution of the arguments of perihelion. This induces variability in the
evolution histories and complicates any attempt to determine the age of the Theobalda family 
using arguments of perihelion. Thus, we selected 13 Theobalda family members which have 
Lyapounov times $T_{lyap}\geq$ $10^{5}$yr and propagated their orbits 20~Myr backwards.
All these members are located at $a_{p}$$\geq$3.183~au. In Fig.~\ref{f:clustering} the average value of $\Delta\Omega$,
for these 13 asteroids, is shown.
Conjunction of nodal longitudes at $\approx$6.2~Myr suggests that the Theobalda family, or at least a
part of the family located at $a_{p}$$\geq$3.183~au, was formed by a catastrophic collision at that time. 
The average $\Delta\Omega$, at $\approx$6.2~Myr, is $\approx$$58^{o}$, much smaller than at any other time. 
This suggests a statistical significance of the $\approx$6.2~Myr event. In this case, however, $\langle\Delta\Omega\rangle$ 
values are substantially more spread at $\approx$6.2~Myr than in the case of Karin cluster ($\langle\Delta\Omega\rangle$ is $\approx$$10^{o}$) 
or Veritas family ($\langle\Delta\Omega\rangle$ $\approx$$40^{o}$).
This is primarily due to two reasons: (1) at least a few MMRs exist in the semi-major axis range from 3.18 to 3.19~au;
thus, despite the present long Lyapounov times of the selected orbits,
these orbits might have experienced periods of chaotic motion in the past; 
and (2) all regular bodies, whose orbits can be accurately tracked back in time,
are small bodies ($\lesssim$5~km) and consequently subject to Yarkovsky thermal force, which, even on this relatively short time scale, 
can produce large enough changes in the semi-major axes, and consequently to affect the secular frequencies in a way that is 
difficult to reconstruct.

In order to estimate how sensitive this result is on the semi-major axis drift due to the Yarkovsky effect, an additional investigation should be carried out. 
As the Yarkovsky induced drift depends on several parameters, we had to decide the values of the parameters characterizing it.
These are asteroid spin axis orientation ($\gamma$), rotational period (P), surface and bulk densities ($\rho$),
surface thermal conductivity (K) and specific heat capacity (C). As Theobalda family members are most likely C-type asteroids,
we have adopted the following values of parameters: $K=0.01-0.5~$~[W\,(m\, K)$^{-1}$], $C=680-1500~$[J\,(K\,kg)$^{-1}$], 
and the same value for surface and bulk density $\rho=1300-1500~$[kg\,m$^{-3}$].
The rotational periods are chosen according to a Gaussian distribution peaked at $P=8~$h, while the distribution of spin axes orientation is assumed to be uniform.
These values are consistent with C-type asteroids \citep{broz06,broz08}.

Next, we made 20 ''yarko'' clones for each of the 13 regular members, by assigning random values of the
parameters, from adopted intervals, to each clone. Then, we integrated\footnote{These integrations were performed using ORBIT9 integrator in the Grid environment \citep{grid09}.} 
the orbits of all clones (260 orbits in total), but accounting not only for gravitational perturbations, but also for Yarkovsky effect. 
The initial orbital elements of the asteroids and planets were the same as in the previous experiment.
Finally, we checked how the value of average $\Delta\Omega$ change with different combinations of clones. We found that the result shown in Fig.~\ref{f:clustering}
is very sensitive to the Yarkovsky induced drift, as expected. In a few cases any significant clustering even disappeared, but in most of the cases we obtained a deeper
minimum. The deepest minimum that we found is related to the clustering within $\approx$$31^{o}$ at about 6.4~Myr ago (Fig.~\ref{f:clustering}), which is still within the error bars obtained
from the integrations without Yarkovsky force. We would like to note here that this high sensitivity of the result on the Yarkovsky parameters
could help us to estimate the rotational periods and spin axis orientations of these 13 asteroids. This can be achieved similarly as was done by \citet{nesbot04} for the Karin cluster members, 
but we reserve this for a future work.

Although, the clustering at about 6.2~Myr within $\approx$$58^{o}$ is the most significant on the time scale of 20~Myr, there is
another clustering at about 15.5~Myr within $\approx$$65^{o}$ (see Fig.~\ref{f:clustering}). As this clustering appears in the more 
distant past where we should expect less tight clustering, it is not possible to rule out its significance.
Also, to use the argument of perihelions is impossible, because the changes in the semi-major axes, caused by Yarkovsky effect, coupled with 
large gradient\footnote{Caused by the proximity of the Theobalda family to the 2/1 resonance with Jupiter.} of secular frequency 
($dg/da$~$\approx$$0.3^{o}$~$yr^{-1}$~au, where $g$ is the longitude of perihelion frequency), erase evolution histories of these angles.
Given that, we believe that, in the case of the Theobalda family, backward integration method is not enough to draw a firm conclusion 
about the age of the family.

\subsection{Chaotic chronology}
\label{ss:mch}

In this section we present results obtained by using MCC in order to estimate the age of 
Theobalda family. This model was successfully applied by \citet{nov09} to estimate ages of Veritas and Lixiaohua asteroid 
families \citep[see also][]{nov09b}. In order to apply MCC 
the family has to be located in the region of the main belt where
diffusion takes place. Also, it is necessary that diffusion is fast enough to cause measurable effects, 
but slow enough so that most of the family members are still forming a robust family structure. 
As our results about diffusion speed suggest, Theobalda family is an excellent example in this respect (see Section~\ref{sss:diffusion}).

The basic steps and the model which we used in our Markov Chain Monte Carlo (MCMC) 
simulations are explained in \citet{nov09}, and thus we will describe these here only briefly.
Our model simulate the evolution of the family in the 3-D space, i.e. proper semi-major axis
$a_{p}$ and two actions $J_{1}$, $J_{2}$ (see Section~\ref{sss:diffusion} for definition of these actions). At the beginning of a simulation the 
\textit{random walkers} are distributed in the region which was presumably occupied by the family members
immediately after the impact event. Then, at each time step $dt$ the random walkers can change their positions in every direction, 
in the 3-D space. The length of the jump in $a_{p}$ is controlled
by Yarkovsky thermal force \citep{farvok99}, while the length of the jumps in 
$J_{1}$ and $J_{2}$ depend on diffusion speed, i.e. on the diffusion coefficients. At the time step 
when 0.3 per cent of random walkers
leave an ellipse\footnote{The ellipse is determined by the present size of the family or, as in this case, by the present size of 
particular part of the family. It should not be confused with equivelocity ellipses shown e.g. in Fig.~\ref{f:dis_aei}.} in the ($J_{1}$, $J_{2}$) 
plane, which corresponds to a 3$\sigma$ confidence
interval of a two-dimensional Gaussian distribution, the simulation stops. The number of time steps
multiplied by the time step $dt$ gives the age of the family.

\subsubsection{Diffusion coefficients}
\label{sss:diffusion}

One of the most important information, which are needed as input for MCMC simulations, are the values of
diffusion coefficients in the region of interest. As was shown by \citet{nov09}, to obtain good estimate of the family age by MCC, 
it is enough to determine diffusion coefficients as a function of proper semi-major axis $a_{p}$. This is our next step.

As well as MCC, the procedure of determination of diffusion coefficients, as the functions of proper semi-major axis, is described in \citet{nov09}.
Let us mention here only its main features and numbers related to this work:
the orbits of $\sim$5,000 fictitious bodies distributed randomly in the same ranges of
osculating orbital elements as the real family members at present, are propagated for 10~Myr; then, the time series of mean elements \citep{MilKne98} 
for all of them are calculated; the mean elements are transformed to actions according to relations\footnote{In these relations $a_{J}$ denotes 
Jupiter's semi-major axis, $e_m$ the mean eccentricity and $I_m$ the mean inclination of the asteroids.} 
$J_{1}\approx\frac{1}{2}\sqrt{\frac{a_{p}}{a_{J}}}e^{2}_{m}$ and
$J_{2}\approx\frac{1}{2}\sqrt{\frac{a_{p}}{a_{J}}}\sin^{2}I_{m}$; next, the family is split in the small cells, in terms of $a_{p}$, 
using a kind of moving-average technique with cell size of $\Delta a_{p} = 5 \times 10^{-4}$~au and step size of $\delta a_{p} = 2 \times 10^{-4}~$au; 
finally, the mean squared displacements $\langle(\Delta J)^{2}\rangle$, for both actions, are calculated, and 
the diffusion coefficients $D(J_{1})$, $D(J_{2})$ for each cell as the least-squares fit slope of the $\langle(\Delta J)^{2}\rangle$(t) curve, are determined.

\begin{figure}
\begin{center}
\includegraphics[width=0.33\textwidth,angle=-90]{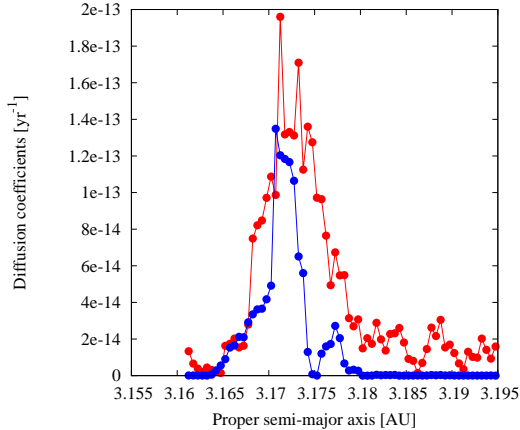}
\end{center}
\caption[]{The values of diffusion coefficients $D(J_{1})$ (red) and $D(J_{2})$ (blue)
in the Theobalda family region, shown here as functions of the proper semi-major axis $a_{p}$. 
Note that $D(J_2)$ is practically zero for $a_{p}\geq$3.18~au.}
\label{f:diff_coef}
\end{figure}

The obtained values of diffusion coefficients $D(J_{1})$ and $D(J_{2})$, in the Theobalda family region are shown in Fig.~\ref{f:diff_coef}.
The fastest diffusion is associated to (5,~-2,~-2) three-body MMR, but the diffusion is very fast in (3,~3,~-2) three-body MMR as well, and these
two chaotic zones seem to be connected. The third chaotic zone, associated to (7,~-7,~-2) resonance, is connected to the first two in terms 
of diffusion in $J_{1}$, but not in terms of diffusion in $J_{2}$. This is in a relatively good agreement with results presented in Section~\ref{ss:dynamics}. 
The diffusion is somewhat faster in $J_{1}$ than in $J_{2}$, and while the local minimum near the center of
(5,~-2,~-2) resonance exist for $D(J_{1})$, similarly as in the case of Veritas family \citep[see][]{nov09}, while there is no such a feature for $D(J_{2})$.
In the region for $a_{p}\geq$3.18~au the values of $D(J_2)$ are practically zero, but there is some diffusion in the $J_{1}$. It should be noted here that 
this might affect age estimation using backward integration method, but not significantly, because the most of 13 asteroids that we used to apply backward 
integration method are located close to $a_{p}=$3.185~au, where both values, $D(J_{1})$ and $D(J_{2})$, are close to zero.

The important general conclusion can be drawn by comparing the values of diffusion coefficients obtained for the region occupied by Veritas family \citep{nov09}
to the values obtained here. Two families stretch over the similar range of the semi-major axis, but members of Theobalda family have somewhat higher inclinations and significantly
higher eccentricities. The estimated diffusion is about one order of magnitude faster in the region occupied by Theobalda family than in the region occupied by
Veritas family. This confirms the fact that chaos is dominant at higher eccentricities.

\subsubsection{Monte Carlo simulations}
\label{ss:mcmc}

Having obtained the values of diffusion coefficients, we are ready to apply MCC to estimate the age of the family.
There are two separate parts of the Theobalda family suitable for application of MCC. These
are bodies inside (5, -2, -2) and (3, 3, -2) three-body MMRs. Following \citet{menios07} who deal with
Veritas family, we called these bodies Group~A (5, -2, -2) and Group~B (3, 3, -2). As the results
about diffusion coefficients confirmed, there exists significant diffusion in both groups. This gives
an unique opportunity to apply MCC to both groups and to obtain two independent age estimates.
A good agreement between these two estimates, as well as with the age derived using backward 
integration method, would suggest a reliable result.

As the present size of the chaotic zone is a critical parameter in our model, we start with
the family as identified by applying HCM for velocity cutoff of $d_{cutoff}=65$~ms$^{-1}$. This
is probably the lowest acceptable value of $d_{cutoff}$ in the case of Theobalda family. 
With this cutoff velocity we identified 30 bodies from Group~A and 16 bodies from Group~B. 
Corresponding  sizes of these groups in $J_{1}$ and $J_{2}$ are: Group~A 
($10.67\pm1.03)\times 10^{-4}$ and ($3.82\pm0.47)\times 10^{-4}$; Group~B 
($10.32\pm1.05)\times 10^{-4}$ and 
($4.00\pm1.06)\times 10^{-4}$.

Using these sizes of two chaotic groups, for each group, we performed 16 sets of MCMC simulations (each set consisting of 100 runs),
by using different number of random walkers $n$ (2000 or 5000), time step $dt$ (from 100~yr to 2000~yr) and for two 
initial sizes of the family which correspond to velocities of $v=35$~ms$^{-1}$ and $v=40$~ms$^{-1}$ (see Fig.~\ref{f:dis_aei}). 
From these simulations we derived the age of family to be 2.5~$\pm$~1.1~Myr (using Group~A bodies) 
and 7.2~$\pm$~3.1~Myr (using Group~B bodies).\footnote{The main source of the error is uncertainty in the determination 
of the present size of the Group~B, due to the small number of members in this Group.} 
The obvious discrepancy between two results needs to be investigated further. The age obtained from Group~B 
is in agreement with what we found using backward integration method, while the age obtained from Group~A suggests 
that the family could be much younger. Also, the discrepancy between ages derived from two different groups,
may be an indication that identification of family members has not been good, i.e. the velocity cut-off of 65~ms$^{-1}$ is too low.

\begin{figure}
\begin{center}
\includegraphics[width=0.33\textwidth,angle=-90]{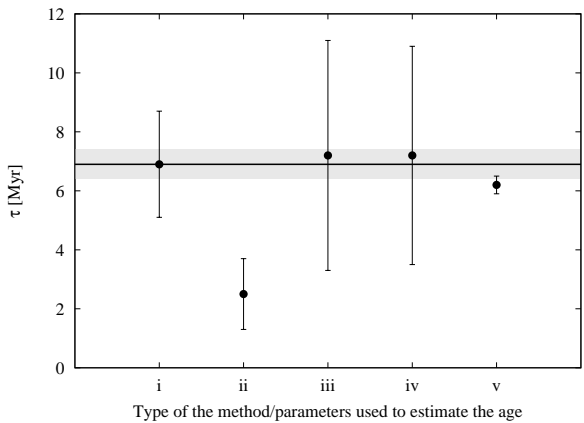}
\end{center}
\caption[]{The age of Theobalda family derived using different methods, asteroids and parameters: (i) MCC applied to Group~A for $d_{cutoff}=85$~ms$^{-1}$;
(ii) MCC applied to Group~A for $d_{cutoff}=65$~ms$^{-1}$; (iii) MCC applied to Group~B for $d_{cutoff}=85$~ms$^{-1}$; (iv) MCC applied to Group~B for 
$d_{cutoff}=65$~ms$^{-1}$; (v) age obtained by backward 
integration method. The bold horizontal line and dashed area correspond to our final estimate of Theobalda family age and its error respectively.}
\label{f:age}
\end{figure} 

Because of that, we repeated all simulations using our nominal family.
For $d_{cutoff}=85$~ms$^{-1}$ we identified 40 bodies from Group~A and 18 bodies from Group~B.
In this case, the corresponding sizes of these groups in $J_{1}$ and $J_{2}$ are: Group~A
($14.86\pm1.24)\times 10^{-4}$ and ($6.03\pm0.65)\times 10^{-4}$; Group~B 
($11.87\pm1.29)\times 10^{-4}$ and ($3.80\pm0.95)\times 10^{-4}$.
From these sizes and the same sets of MCMC simulations as in the previous case, we derived the age
of Theobalda family to be 6.9~$\pm$~1.8~Myr (using Group~A bodies) and 7.2~$\pm$~3.0~Myr (using Group~B bodies).

Now, the agreement between two results is very good, and also, both results agree quite well with the age obtained by 
backward integration method.\footnote{The good agreement between ages obtained applying MCC to groups A i B, is one of the reasons why
we adopted value of $d_{cutoff}=85$~ms$^{-1}$ to identify nominal family. It should be noted here that identification of resonant family members 
is not straightforward. Too small cut-off, on one hand,
may prevents identification some of the real family members. On the other hand, the large cut-off could associate some interlopers with family.
This may be one of the reasons for the variations of age estimates obtained by MCC for two different cut-off velocities.}
This, in our opinion, is a very strong indication that Theobalda family was formed about 7~Myr ago.

The fact, that four out of five, different age estimates, agree well, is the reason why we reject the age of 2.5~$\pm$~1.1~Myr, 
derived using Group~A ($d_{cutoff}=65$~ms$^{-1}$), as a possible solution. Thus, in order to obtain our final estimate of the age of Theobalda family, 
we use four results which are in a good agreement (see Fig.~\ref{f:age}). The values of the mean ($\mu$) and standard deviation ($\sigma$) of non-overlapping 
sub-samples, of the same size, can be calculated as:

\begin{equation}
\mu = \frac{\Sigma_{i=1}^{m}\mu_{i}}{m}
\label{eq:mean}
\end{equation}

\begin{equation}
 \sigma=\sqrt{\frac{\Sigma_{i=1}^{m} ((r-1)\sigma_{i}^{2}+r\mu_{i}^{2}) - mr\mu^{2}}{mr-1}}
\label{eq:std}
\end{equation}
where $m$ is the number of sub-samples, $r$ is the size of each sub-sample (100 in our case), $\mu_{i}$ is the mean of $i$-th sub-sample, and
$\sigma_{i}$ is the standard deviation of $i$-th sub-sample.
Using Eqs.~\ref{eq:mean} and \ref{eq:std} we obtain the final age estimate, Theobalda asteroid family is 6.9~$\pm$~2.3~Myr old.

\section{Summary, Discussions and Conclusions}
\label{s:conclusions}

We have presented here a detailed study of Theobalda asteroid family. 
We found that family now consists of 128 members. By analyzing SFD of the identified family members we were able
to infer diameter of the parent body to be $D_{PB}$~$\approx78\pm9$~km. However, this estimate is based on 
the assumption that all family members have the same albedo as the largest family member, asteroid (778)~Theobalda. 
In order to obtain better estimate, the albedos of as many as possible family members are desirable. 
Ongoing projects, such as Wide-Field Infrared Survey Explorer (WISE) should improve situation significantly in this respect.

The most, but not all, of Theobalda family members move on chaotic orbits, thus, giving rise to 
the significant chaotic diffusion which has been changing the kinematical structure of the family over time.
The study of dynamical characteristics, in the region occupied by the family, showed that three three-body MMRs
are the most efficient in shaping the family. These are (3,~3,~-2), (5,~-2,~-2) and (7,~-7,~-2) resonances, 
and they are connected in this high eccentricity region, allowing bodies to switch from one resonance to another.

The fact that some of the family members have stable orbits was the reason why we were able to apply backward integration
method to estimate the age of the family. On the other hand, presence of the chaos in the region occupied by the family, 
allows to use MCC in order to estimate the age. Using both methods, and combining the results, we found the age of Theobalda family to be
6.9~$\pm$~2.3~Myr. Given the very good agreement between results obtained with different methods as well as when applied to different
groups, we believe this estimate is very robust. Thus, this is another family younger than 10~Myr. This result has several important
implications, and some of them we mention bellow.

The young asteroid families are also known to be source of solar system dust bands \citep[see e.g.][]{grun85,nes2003}.
The origin of three main dust bands is known, and they correspond to Karin, Veritas and Beagle family \citep{nes2006b,nes2008}.
Also, the very young Emilkowalski family is the most probably source of incomplete dust band at $17^{o}$ \citep{espy2009}.
On the other hand, the origin of some less prominent bands, such as so called M/N dust band, is still not quite clear.
The Theobalda family's young age, and its
proper inclination of $I_{p}$~$\approx14^{o}.3$ suggest that it might be a possible source 
of M/N dust band ($I_{p}$~$\approx15^{o}$) \citep{sykes90}.
On the other hand, this dust band was linked to (170) Maria asteroid family \citep{reach97}, and 
more recently to (1521) Seinajoki cluster \citep{nes2003}.
In any case, dust band produced by such a young family, as Theobalda, should be observable. 
The size of its parent body also suggests that it should produce a prominent dust band. 
If this is not M/N dust band, then there must be another dust band which can be linked to this family. 
Alternatively, it should be explained why and how this dust band has disappeared.

Theobalda asteroid family is located very close to the region where three (out of four)
so-called \textit{main belt comets} (MBCs)\footnote{The MBCs are bodies with asteroid-like
dynamical properties but comet like physical properties \citep{hsieh2004}.
These are dynamically ordinary main-belt asteroids on which, probably, subsurface 
ice has recently been exposed e.g. because of a collision.} have been discovered 
\citep[see e.g.][]{jewitt2009}. As was suggested by \citet{hsieh2009}, it is possible 
that this kind of bodies can be found among the members of other young families,
probably many of which waiting to be discovered. Being young and dominated by C-type asteroids,
we believe Theobalda family is very good place to start.

\section*{Acknowledgements}
I am grateful to Zoran Kne\v zevi\' c and Rade Pavlovi\' c for their useful suggestions on the manuscript. 
I also would like to thank David Nesvorn\' y, the referee, for his useful comments and suggestions
that help me to improve this article.
This work has been supported by the Ministry of Science and Technological Development
of the Republic of Serbia (Project No 146004 "Dynamics of
Celestial Bodies, Systems and Populations").

\end{document}